\documentclass[preprint,12pt]{elsarticle}
\usepackage{amsmath}
\usepackage{amssymb}
\usepackage{graphicx}
\usepackage{color}

\newcommand{\cov}{\mbox{cov}}

\def\bfg{\mbox{\boldmath{$g$}}}
\def\bfx{\mbox{\boldmath{$x$}}}
\def\bfy{\mbox{\boldmath{$y$}}}
\def\bfX{\mbox{\boldmath{$X$}}}
\def\bfY{\mbox{\boldmath{$Y$}}}
\def\bflambda{\mbox{\boldmath{$\lambda$}}}
\def\bftheta{\mbox{\boldmath{$\theta$}}}

\def\hbftheta{\hat{\bftheta}}

\def\bfdelta{\mbox{\boldmath{$\delta$}}}
\def\bfw{\mbox{\boldmath{$w$}}}
\def\bfq{\mbox{\boldmath{$q$}}}

\def\bfU{\mbox{\boldmath{$U$}}}
\def\bfo{\mbox{\boldmath{$o$}}}
\def\bfO{\mbox{\boldmath{$O$}}}

\def\hbflambda{\hat{\bflambda}}

\newcommand{\zero}{{\mbox{\boldmath$0$}}}
\def\bfone{\mathbf{1}}
\newcommand{\non}{\nonumber}
\newcommand{\be}{\begin{equation}}
\newcommand{\ee}{\end{equation}}
\newcommand{\bea}{\begin{eqnarray}}
\newcommand{\eea}{\end{eqnarray}}
\newcommand{\beaa}{\begin{eqnarray*}}
\newcommand{\eeaa}{\end{eqnarray*}}
\newcommand{\ep}{\varepsilon}
\newcommand{\Int}{{\rm{int}}}
\newtheorem{thm}{Theorem}

  \usepackage{amssymb}

\begin{document}

\begin{frontmatter}

\title{ Data Fusion Using Robust Empirical Likelihood Inference  }

\author{  Hsiao-Hsuan Wang , Yuehua Wu,  Yuejiao Fu and  Xiaogang Wang\\ \vskip 0.1in Department of Mathematics and Statistics, York University,  Canada. }

\address{}

\begin{abstract}
The authors propose a robust semi-parametric empirical likelihood
method to integrate all available information from  multiple samples with a common center of measurements.
 Two different sets of estimating equations are used to improve the classical likelihood inference
 on the measurement center.
 The proposed  method   does not require the knowledge
 of the functional forms of the probability density functions of related populations.
 The advantages  of the proposed method were
demonstrated through the extensive simulation studies by comparing mean squared error,
coverage probabilities and average length of confidence intervals with those from the classical likelihood method.  
Simulation  results
suggest that  our  approach provides more informative and efficient inference than the conventional maximum likelihood
estimator when certain structural relationships exist  among the parameters for these relevant samples.

\end{abstract}

\begin{keyword}


Data Fusion; Empirical likelihood; Robust estimation; Multiple samples
\end{keyword}

\end{frontmatter}


\section{Introduction}
\label{intro}

A common problem in clinical trials and medical research is how to
accurately and
efficiently estimate parameters of interest when the current sample size is
small due to   cost   and time constraints.  Usually
there might exist certain surrogate     populations with low sampling cost that could provide  relevant
information  for the population of direct inferential interest.
In this article, we propose a robust semi-parametric method to integrate related information from different
sources to improve the classical likelihood method.

The classical likelihood approach is arguably the most widely
used method in statistical inference.  It has been routinely applied in almost all
the statistical applications.  Despite the
great success and excellent asymptotic properties, the
classical likelihood has  limitations associated with making inference for small sample sizes.
Consider a thought experiment as follows. Suppose that a random
experiment is to toss a coin twice. The
parameter of interest, denoted as $\theta_1$,
is the probability of turning up head for this
coin. The maximum likelihood estimator (MLE) of
$\theta_1$ is denoted as $\hat{\theta}_1$.  If
the coin is a fair one, the MLE will obtain the
following $P(\hat{\theta}_1= 0 \mbox{ or }
1)=1/2$ and $P(\hat{\theta}_1 = 1/2) = 1/2$.
Thus, one would have $50\%$ chance to make a
nonsensical decision by using the MLE when the
sample size is only two. In addition, suppose that for
some reason we cannot use this coin any more
but we can flip another coin instead. In this
situation, the classical likelihood approach
would not consider the second experiment since
it comes from a different population unless a functional relationship between the two
parameters is known.  If the
second population is related to the first one  due to some unknown link between these two parameters,  
one should be able to utilize this connection and make better statistical inference.

Different statistical methodologies have been proposed
in the literature to integrate information from different sources (or populations) in a  very general setting, see Wang, van Eeden
and Zidek (2004), Fu, Wang and Wu (2009) and referees therein.
Most of these methods, however, face the challenge of
accurately validating or evaluating  the relevance of all related information
to handle the possibility of introducing a significant bias or contaminating the current sample. In other words, the  magnitude of integration
must be controlled carefully and in addition likelihood weights must be chosen judiciously in order to achieve any
desired improvement in statistical inference.
We propose to tackle this difficult problem using a robust semi-parametric empirical likelihood
method to gain more accurate and robust inferential results.

Empirical likelihood, which was first introduced by Owen (1988), is
a nonparametric method of inference based on a data-driven
likelihood ratio function.  It allows the statistician to employ
likelihood methods, without specifying a parametric model for the
data.  It enjoys both the flexibility of nonparametric methods and
the efficiency of parametric likelihood.  As shown in Qin and
Lawless (1994), empirical likelihood is a prominent efficient tool
in estimating parameters by incorporating estimating equations into
constrained maximization of the empirical likelihood function.
In the problem we consider, the relevant information from different sources could be
used by incorporating extra set of estimating equations in the empirical likelihood framework.

To obtain robust estimates, we use median as an estimate of center
instead of mean.  We propose here using two different kinds of
estimating equations; one uses median and the other one uses a 
smoothed version of median.  The smoothing technique is the one proposed
by Shi and Lau (1999) to improve the coverage accuracy. Our
method can be easily generalized to multiple samples with relevant
information. Without loss of generality, we consider data with two
populations.

The rest of the paper is organized as follows. The methodology
framework, the proposed empirical likelihood approach, and its
theoretical properties are presented in Section \ref{s: RWEL
Method}. Results of simulation studies demonstrating the empirical
performance are provided in
Section 3. Conclusion and some discussion
are provided in Section \ref{s: RWEL Conclusion}.

\section{Methodology}
\label{s: RWEL Method}

Suppose there are two groups of data from different population but
sharing the same parameter of interest.  Assume that
\[
\bfx_1,\ldots,\bfx_{n_1}\sim f(\bfx,\bftheta).
\]
The second group of data $\bfy_1,\ldots,\bfy_{n_2}$ might be
different from the first population, and
\[
\bfy_1,\ldots,\bfy_{n_2}\sim h(\bfy,\bftheta).
\]
Our goal is to estimate $\bftheta$ by using both samples.
Directly using the log-likelihood
\[
\sum_{i=1}^{n_1} \log f(\bfx_i,\bftheta)+\sum_{i=1}^{n_2} \log
h(\bfy_i,\bftheta),
\]
we might get a biased estimation due to the difference between the two
populations.

We propose a semi-parametric empirical likelihood method which
only requires the independence of these two samples. To
combine the second sample with the first one, we use the following semi-parametric empirical likelihood
\[
\ell=\sum_{i=1}^{n_1} \log f(\bfx_i,\bftheta)+\sum_{i=1}^{n_2} \log
n_2p_i,
\]
where
\[
p_i\geq 0, \hspace{1em} \sum_{i=1}^{n_2}p_i=1, \hspace{1em}
\sum_{i=1}^{n_2}p_i\bfg({\bfy}_i,\bftheta)={\bf 0},
\]
and $\bfg({\bfy}_i,\bftheta)$ is an estimating function. From the
empirical likelihood theory, we know $p_i$ is maximized by
$1/\{n_2\left[1+\bflambda^T\bfg({\bfy}_i,\bftheta)\right]\}$, where
$\bflambda$ is the Lagrange multiplier.  We can rewrite the log
likelihood function as
\begin{equation} \label{RWEL: eqn main }
\ell(\bftheta)=\sum_{i=1}^{n_1} \log
f(\bfx_i,\bftheta)-\sum_{i=1}^{n_2}
\log\big[1+\bflambda^T\bfg({\bfy}_i,\bftheta)\big],
\end{equation}
and
\[\hat{\bftheta}=\arg\max_{\bftheta} \ell(\bftheta).\]
We call $\hat\bftheta$ the robust semi-parametric empirical
likelihood estimate (RSPELE).

The advantage of the log profile likelihood function is that it does
not depend on the likelihood weights which could be difficult to choose.
In our propose method, we do not require  that the probability density function  of
the second population is identical to the first
population.  By using the empirical likelihood method, we do not even need to
specify the functional form of the underlying distribution of the
second population. Therefore,  we can gain robust estimates in the
sense that   model mis-specification problem is avoided. Consequently, our method can be employed
in a relatively wide range of applications when the functional form of the probability density function is not known.

In the following, the theoretical properties of the proposed RSPELE estimator will be presented.
For clarity, all
proofs are postponed to the Appendix.
Theorem \ref{WEL thm:
consistency} below shows that under some regularity conditions, the RSPELE estimator
$\hbftheta$ is consistent to $\bftheta_0$.

\begin{thm} \label{WEL thm: consistency}
Let $\bfx_1,\ldots,\bfx_{n_1}$ be i.i.d.\ from $f(\bfx,\bftheta)$
and $\bfy_1,\ldots,\bfy_{n_2}$ be i.i.d.\ from an unknown
distribution. Assume that $f(\bfx,\bftheta)$ satisfies the
regularity conditions given in Shao (2003) on the normality of the
maximum likelihood estimator in parametric models.  Let $\bftheta_0$
be the true parameter.  We further assume that
\begin{description}
  \item[(A1)] There exists a matrix $\Psi>0$ such that
\[E\big[\bfg({\bfy}_i,\bftheta)\big]=\Psi(\bftheta-\bftheta_0)
+o(\|\bftheta-\bftheta_0\|)\] as $\bftheta-\bftheta_0\to {\bf 0}$.
  \item[(A2)] $\zeta(\bftheta)\equiv
E\big[\|\bfg({\bfy}_i,\bftheta)-\bfg({\bfy}_i,\bftheta_0)\|^2\big]$
exists when $\|\bftheta-\bftheta_0\|$ is sufficiently small and is
continuous at $\bftheta_0$
  \item[(A3)] $E\big[\bfg({\bfy}_i,\bftheta_0)\bfg^T({\bfy}_i,\bftheta_0)\big]>0$.
\end{description}
Then, it follows that $\hbftheta\to\bftheta_0$ in probability in the
neighborhood of $\bftheta_0$ such that $\{\bftheta:
||\bftheta-\bftheta_0||\leq n^{-1/2}\}$.
\end{thm}

The asymptotic distribution of the
$\bftheta$ and $\bflambda$ is shown in Theorem \ref{WEL thm: asymptotic}.

\begin{thm} \label{WEL thm: asymptotic}
In addition to the conditions of Theorem \ref{WEL thm: consistency},
we assume that $\displaystyle\frac{n_1}{n}\to b$, where $b$ is a
constant.  We also assume that
$\displaystyle\frac{\partial\bfg({\bfy},\bftheta)}{\partial\bftheta}$
exists with probability one and the set of its discontinuity points
has zero probability.  Then \beaa \sqrt{n}(\hbftheta-\bftheta_0)
&\stackrel{d}{\longrightarrow}&N({\bf 0},S_1),\\
\sqrt{n}\hbflambda &\stackrel{d}{\longrightarrow}&N({\bf
0},S_2),\eeaa where \bea \non
S_0&=&bI(\bftheta_0)+(1-b)E\left(\displaystyle\frac{\partial
\bfg}{\partial\bftheta^T}\right)^T \Phi^{-1}
E\left(\displaystyle\frac{\partial
\bfg}{\partial\bftheta^T}\right),\\
\nonumber
S_1&=&bS_0^{-1}I(\bftheta_0)S_0^{-1}+(1-b)S_0^{-1}E\left(\displaystyle\frac{\partial
\bfg}{\partial\bftheta^T}\right)^T \Phi^{-1}
 E\left(\displaystyle\frac{\partial
\bfg}{\partial\bftheta^T}\right)S_0^{-1},
\\ \non
S_2&=&b\Phi^{-1} E\left(\displaystyle\frac{\partial
\bfg}{\partial\bftheta^T}\right)S_0^{-1}I(\bftheta_0)S_0^{-1}
E\left(\displaystyle\frac{\partial
\bfg}{\partial\bftheta^T}\right)^T\Phi^{-1}\\
\non && +\left[-\Phi^{-1} + (1-b)
\Phi^{-1}E\left(\displaystyle\frac{\partial
\bfg}{\partial\bftheta^T}\right)S_0^{-1}E\left(\displaystyle\frac{\partial
\bfg}{\partial\bftheta^T}\right)^T\Phi^{-1} \right] \\ \non
&&\times\Phi\left[-\frac{1}{1-b}\Phi^{-1} +
\Phi^{-1}E\left(\displaystyle\frac{\partial
\bfg}{\partial\bftheta^T}\right)S_0^{-1}E\left(\displaystyle\frac{\partial
\bfg}{\partial\bftheta^T}\right)^T\Phi^{-1} \right],
\\
\Phi&=& E\bfg\bfg^T, \eea and $I(\bftheta_0)$ is the Fisher
information about $\bftheta_0$ contained in $\bfX$.
\end{thm}

The asymptotic distribution of
$2[\ell(\hbftheta,\hbflambda)-\ell(\bftheta_0,\bflambda_0)]$ is
given in Theorem \ref{WEL thm: hypothesis}.

\begin{thm} \label{WEL thm: hypothesis}
Assume that the assumptions made in Theorem \ref{WEL thm:
asymptotic} hold.  The limiting distribution of
$2\ell(\hbftheta,\hbflambda)-2\ell(\bftheta_0,\bflambda_0)$ is the
same as the distribution of
\[\left(\begin{array}{c}U_1\\
U_2\end{array}\right)^TV^{-1}\begin{pmatrix}
            S_0 & (1-b)E\left(\displaystyle\frac{\partial
            \bfg}{\partial\bftheta^T}\right)^T \\
           (1-b)E\left(\displaystyle\frac{\partial
           \bfg}{\partial\bftheta^T}\right) & 0 \\
          \end{pmatrix}V^{-1}
           \left(\begin{array}{c}U_1\\
U_2\end{array}\right)\] under $H_0: \bftheta=\bftheta_0$, where
$\bfU_1$ is independent of $\bfU_2$,
\[
\bfU_1\sim N({\bf 0}, bI(\bftheta_0)),\quad \bfU_2\sim N\Big({\bf
0},\;(1-b)E\big[\bfg(\bfY,\bftheta_0)\bfg^T(\bfY,\bftheta_0)\big]\Big),
\] and \[
V^{-1}=\begin{pmatrix}
            S_0^{-1} & S_0^{-1}E\left(\displaystyle\frac{\partial
            \bfg}{\partial\bftheta^T}\right)^T\Phi^{-1} \\
           \Phi^{-1}E\left(\displaystyle\frac{\partial
           \bfg}{\partial\bftheta^T}\right)S_0^{-1} & V^{(22)} \\
          \end{pmatrix},
\] where $V^{(22)}=-\frac{1}{1-b}\Phi^{-1} +
\Phi^{-1}E\left(\displaystyle\frac{\partial
\bfg}{\partial\bftheta^T}\right)S_0^{-1}E\left(\displaystyle\frac{\partial
\bfg}{\partial\bftheta^T}\right)^T\Phi^{-1}$.
\end{thm}
We note that other test statistics, for example, a test
statistic based on Theorem \ref{WEL thm: asymptotic}, may also be
used.

Estimating equations provide a very flexible way to specify how the
parameters of a statistical model should be estimated. They serve as
constraints in maximizing the empirical likelihood. Qin and Lawless
(1994) showed that the empirical likelihood method is an efficient tool
for point estimation through estimating equations. In this section,
we consider two different kinds of estimating equations using the
information of median, since median is robust with respect to the
outliers, one may use
\[
\bfg_1({\bfy}_i,\bftheta)=\bfone(\bfy_i\leq m(\bftheta))-1/2
\]
as estimating function based on the second group data, where $
m(\bftheta)$ is the median of $f(\bfx,\bftheta)$ and $\bfone(\cdot)$
is the usual indicator function.  It is easy to verify that
$E(\bfg_1({\bfy}_i,\bftheta))=0$.

Due to the discontinuity of $\bfg_1$, we may use the smoothed
version of the constraint  which was motivated by Shi and Lau
(1999). First of all, we define the estimating equation for the
smoothed empirical likelihood.  In general, let $\kappa$ be the
$r^{th}$-order kernel (Shi and Lau, 1999), such that
\begin{eqnarray}\renewcommand{\arraystretch}{0.9} \nonumber
\int u^j\kappa(u)du = \left\{\begin{array}{cl}
                1,  & \mbox{ if } j=0, \\
                0,  & \mbox{ if } 1\leq j\leq r-1.\\
                c_r,& \mbox{ if } j=r,
              \end{array}
              \right.
\end{eqnarray}
where $r$ is a positive integer.  Define $\psi(u)=\int
\kappa(t)I(t<u)dt.$ For any $h>0$, let $\psi_h(u)=\psi(u/h)$ where
$h$ is called the smoothing parameter. The kernel $\kappa$ is a
symmetric probability density with bounded and compact support.  Let
the estimating function $g_2(y,\theta,h)=\psi_h(m(\theta)-y)-1/2$.
Therefore, $g_2(y,\theta,h)$ is continuous with respect to $y$, but
it is not a fixed function as the smoothing parameter varies. See
Shi and Lau (1999) for details.  In addition, by using the
arguments similar to those stated in the Theorems \ref{WEL thm:
consistency} to \ref{WEL thm: hypothesis} and Shi and Lau (1999), we
may get the similar asymptotic results.

\section{Numerical Experiments}

\subsection{Data Fusion with Conventional Empirical Likelihood}
\label{s: RWEL Simulation} 
Simulation studies are carried out by performing data fusion when two samples are available.
The first sample
$\bfX=(X_1,\ldots,X_{n_1})$ is generated from standard normal
distribution and the second sample $\bfY=(Y_1,\ldots,Y_{n_2})$ is
generated from normal, double exponential, or $t$-distribution respectively. The
sample size of first sample, $n_1$, is 10 and for the second sample,
the sample size, $n_2$, varies from 10, 20 to 30. 

First of all we
use the median constraint, so the log likelihood function of the
simulation model is
\[
l(\mu)=-\sum_{i=1}^{n_1}\left\{\log\sqrt{2\pi
s_1^2}+\frac{(x_i-\mu)^2}{2s_1^2}\right\}-\sum_{i=1}^{n_2}\log\left\{1+\lambda
[I(y_i\leq\mu)-1/2]\right\},
\]
where $s_1^2$ is the MLE variance,
$s_1^2=\frac{1}{n_1}\sum_{i=1}^{n_1}(x_i-\bar{x})^2$. 

We present the
mean square error (MSE) ratio of RSPELE to MLE based on 1,000
replications in Table \ref{RWEL table: normal}. The simulation results show that RSPELE performs well except in the
situation when t  second population
is normally distributed  with large variation  as the first
one.  When the second sample size   is
increasing, the RSPELE becomes more accurate.  Moreover, we have smaller
MSE of RSPELE when the data of the second population is more
concentrated around the center, for example,   the double
exponential distribution.

%
%
%
%
%

\begin{table} [h]
  \centering
  \caption{MSE Ratio of RSPELE to MLE based on 1,000 replications.}
  \renewcommand{\arraystretch}{0.75}
  \renewcommand{\tabcolsep}{0.4mm}
  \footnotesize \centering
  \begin{tabular}{ccccccccccc}
    \hline\hline
    \multicolumn{11}{c}{\small\centering{\itshape Distribution of sample 2}} \\
    $n_2$ & $N(0,1)$ & $N(0,1.25)$ & $N(0,1.5)$ & $N(0,2)$ & $N(0,3)$ & \hspace{1.5em}$t_3$\hspace{1.5em}
     & \hspace{1.5em}$t_5$\hspace{1.5em} & $DE(0,0.5)$  & $DE(0,1)$ & $DE(0,1.5)$ \\
    \cline{2-11}\\
    10 & 0.776 &  0.857 &  0.929 &  1.017 &  1.028 &  0.787 &  0.747 &  0.431 &  0.717 &  0.875 \\\\
    20 & 0.569 &  0.693 &  0.769 &  0.871 &  0.956 &  0.621 &  0.580 &  0.209 &  0.515 &  0.705 \\\\
    30 & 0.453 &  0.559 &  0.661 &  0.789 &  0.926 &  0.462 &  0.467 &  0.137 &  0.387 &  0.582 \\
    \hline\hline
  \end{tabular}
  \label{RWEL table: normal}
\end{table}

\subsection{Smoothed Empirical Likelihood}

In this section we demonstrate the smoothed version of the
estimating equations.  The kernel we chose is the same as the one
used in Shi and Lau (1999),
\begin{eqnarray} \renewcommand{\arraystretch}{0.9} \nonumber
\kappa(u) = \left\{\begin{array}{cl}
                \frac{3}{4\sqrt5}(1-\frac{1}{5}u^2), & \mbox{ if } \|u\|\leq\sqrt{5}, \\
                0 & \mbox{otherwise}.
              \end{array}
              \right.
\end{eqnarray}
The simulation model is identical to the first experiment. Four values of the
smoothed parameter are used which are $n_2$ to
the power of -1, -3/4, -1/2 and -1/4. The log
likelihood function of the simulation model is
\[
l(\mu)=-\sum_{i=1}^{n_1}\left\{\log\sqrt{2\pi
s_1^2}+\frac{(x_i-\mu)^2}{2s_1^2}\right\}-\sum_{i=1}^{n_2}\log\left\{1+\lambda
g_2(y_i,\theta,h)\right\},
\]
where
$s_1^2=\frac{1}{n_1}\sum_{i=1}^{n_1}(x_i-\bar{x})^2$
is the MLE variance. 

We provide the MSE ratio of
RSPELE to MLE based on 1,000 replications in
Table \ref{RWEL table: Smooth with h=n_2}.
Results of the smoothed version are slightly better than the
results of the median version, no matter which smoothing parameter
is chosen.  When the underlying distribution of the second
population is not the same as the first population, the RSPELE
estimate performs better than the MLE.  When the sample size of the
second population is increasing, the RSPELE estimate is more
accurate.

\subsection{Confidence Intervals}

In this subsection, we construct the confidence
interval for the median by bootstrapping.  In
this simulation study, the first sample $X$ is
generated from standard normal distribution and
the second sample $Y$ is generated from normal,
double exponential, or $t$-distribution. The
sample size of $X$ is 10 and of $Y$ varies from
10, 20, to 30.  The size of the bootstrapped sample
is 200 and the number of iterations is set to be 1,000.  First
of all, we use the median estimating equation
and record the coverage probabilities and the
simulated average confidence interval lengths
(AL) in Table \ref{RWEL table: CP RSPEL Median}
for nominal levels of 80, 90, 95, and 99
percent.  The coverage probabilities and the AL
of using the smoothed version of the estimating
equation are recorded in Tables \ref{RWEL
table: CP RSPEL n_2^{-1}} and \ref{RWEL table:
CP RSPEL n_2^{-1/2}} with different smoothing
parameters which are $n_2$ to the power of -1
and -1/2. We report the results of MLE in Table
\ref{RWEL table: CP MLE}. Since   the
results of MLE do not depend on the second
population, we further compare the
coverage probabilities and AL as in Tables
\ref{RWEL table: CP RSPEL Median} to \ref{RWEL
table: CP RSPEL n_2^{-1/2}} with Table
\ref{RWEL table: CP MLE}.

The results of smoothed version are better than median version in
terms of the coverage probabilities.  The coverage probabilities of
RSPELE and MLE are very close but the confidence intervals of RSPELE
are about 10\% narrower than of MLE.  The results of RSPELE when the
underlying distribution of the second population is either $t$ or
double exponential distribution are better than the results of
RSPELE when underlying distribution is normal distribution.  That is
because normal distribution is flatter than t and double
exponential. Consequently, if the second population provides a good
information about the center we can use it to get   better estimates.

\section{Discussions}
\label{s: RWEL Conclusion}

In this paper, we propose a robust semi-parametric empirical
likelihood  in a multiple-sample model with common
measurement of center. We use two different kind of estimating
equations of information about the median.  
Simulation studies have shown that  the second population could provide
very  useful information on the
parameter of interest by comparing the performance of various commonly used
measures for evaluations.

\vskip 0.2in {\bf Acknowledgements:} We would like to thank Dr. Jing
Qin for his generous and valuable comments and suggestions for this
article.


\section*{Appendix: Proofs}
\label{s: RWEL appedix}
\subsection*{Proof of Theorem \ref{WEL thm: consistency}}

We rewrite the equation (\ref{RWEL: eqn main }) as
$\ell(\bftheta)=\ell_1(\bftheta)+\ell_2(\bftheta)$, where
$\ell_1(\bftheta)= \sum_{i=1}^{n_1}\log f(\bfx_i,\bftheta)$ and
$\ell_2(\bftheta)=-\sum_{i=1}^{n_2}\log
[1+\hbflambda^T\bfg(\bfy_i,\bftheta)]$. We denote
 \begin{equation} \nonumber
 \begin{array} {l}
   N(\bftheta_0) \mbox{ be the neighborhood of } \bftheta_0 \mbox{ such
 that }\{\bftheta:\ \|\bftheta-\bftheta_0\|\leq n^{-1/2}\};\\
    \partial N(\bftheta_0)$ be the boundary of $
  N(\bftheta_0)$, i.e. all $\bftheta_*$ such that $
  \|\bftheta_*-\bftheta_0\|= n^{-1/2}; \\
    N_{\Int}(\bftheta_0)$ be the neighborhood of $
  \bftheta_0$ such that $\{\bftheta:\ \|\bftheta-\bftheta_0\|<
  n^{-1/2}\}; \\
   n=n_1+n_2.
   \end{array}
 \end{equation}

\textbf{Case 1.} $0<b_1<\displaystyle\frac{n_1}{n_2}<b_2<\infty$,
where $b_1$ and $b_2$ are two constants.

In view of the proof of Theorem 4.17 of Shao (2003), it follows that
for any $\ep>0$,  \be P\Big[\ell_1(\bftheta_*)-\ell_1(\bftheta_0)<0
\quad \mbox{for all } \bftheta_*\in \partial N(\bftheta_0)\Big]\geq
1-\ep, \label{l1} \ee for large $n_1$.

By Assumptions (A1)-(A3), applying a similar approach as in Owen
(2001), it can be shown that $\bflambda=\bfO(n^{1/2})$.  In light of
Bai, Rao and Wu (1992), Qin and Lawless (1994), and under the
assumptions (A1)-(A3), it follows that for any $\ep>0$, \be
P\Big[\ell_2(\bftheta_*)-\ell_2(\bftheta_0)<0 \quad \mbox{for all }
\bftheta_*\in \partial N(\bftheta_0)\Big]\geq 1-\ep, \label{l2} \ee
for large $n_2$.

Combining (\ref{l1}) with (\ref{l2}), we have for any $\ep>0$,
\[
P\Big[\ell(\bftheta_*)-\ell(\bftheta_0)<0 \quad \mbox{for all }
\bftheta_*\in \partial N(\bftheta_0)\Big]\geq 1-\ep,\]for large $n$.
Therefore, there exists $\hbftheta\in N_{\Int}(\bftheta_0)$ such
that
\begin{eqnarray}
&&\frac{\partial\ell(\hbftheta)}{\partial\bftheta}={\bf 0}
\quad\mbox{and}\quad\hbftheta= \arg\max_{ \hbftheta\in
N_{\Int}(\bftheta_0)}\ell(\bftheta)\non.
          \end{eqnarray}
 By the definition of $N(\bftheta_0)$, it follows that
$\hbftheta\to\bftheta_0$ in probability.

Case 2. $\displaystyle\frac{n_1}{n_2}\to \infty$ or
$\displaystyle\frac{n_1}{n_2}\to 0$.

The consistency of $\hbftheta$ can be shown similarly. The details
are omitted.\hspace{1cm} $\square$
\subsection*{Proof of Theorem \ref{WEL thm: asymptotic}}


We denote

$
\bfw_{n_1}(\bftheta)\equiv\displaystyle\frac{\partial\ell_1(\bftheta)}{\partial\bftheta}
=\displaystyle\sum_{i=1}^{n_1}\frac{\partial\log
f(\bfx_i,\bftheta)}{\partial\bftheta}; $

$
  \bfq_{1n_2}(\bftheta,\bflambda)\equiv\displaystyle\frac{\partial\ell_2(\bftheta)}{\partial\bftheta}
=-\displaystyle\sum_{i=1}^{n_2}\frac{1}{1+\bflambda^T\bfg({\bfy}_i,
\bftheta)}\frac{\partial\bfg({\bfy}_i,\bftheta)}{\partial\bftheta^T}\bflambda;
$

$
  \bfq_{2n_2}(\bftheta,\bflambda)=\displaystyle\sum_{i=1}^{n_2}\frac{1}{1+\bflambda^T
\bfg({\bfy}_i,\bftheta)}\bfg({\bfy}_i,\bftheta). $

\noindent Since
\[\frac{\partial\ell(\hbftheta)}{\partial\bftheta}={\bf
0}\quad\mbox{and}\quad\bfq_{2n_2}(\hbftheta,\hbflambda)={\bf 0},\]
by applying Taylor's expansion, it follows that \bea {\bf 0}&=&
\bfw_{{n_1}}(\hbftheta)+\bfq_{1n_2}(\hbftheta,\hbflambda)\non\\
&=& \bfw_{{n_1}}(\bftheta_0)+\frac{\partial
\bfw_{{n_1}}(\bftheta_0)}{\partial\bftheta^T}(\hbftheta-\bftheta_0)+\frac{\partial
\bfq_{1n_2}(\bftheta_0,{\bf
0})}{\partial\bflambda^T}\hbflambda+o_p(\bfdelta_n),\non \eea and
\bea
{\bf 0}&=&\bfq_{2n_2}(\hbftheta,\hbflambda)\non\\
\non\\
&=&\bfq_{2n_2}(\bftheta_0,{\bf 0})+\frac{\partial
\bfq_{2n_2}(\bftheta_0,{\bf
0})}{\partial\bftheta^T}(\hbftheta-\bftheta_0) + \frac{\partial
\bfq_{2n_2}(\bftheta_0,{\bf
0})}{\partial\bflambda^T}\hbflambda+o_p(\bfdelta_n), \non \eea where
$\bfdelta_{n}=\|\hbftheta-\bftheta_0\|+\|\hbflambda\|$. It is noted
that \beaa \frac{\partial
\bfw_{n_1}(\bftheta_0)}{\partial\bftheta^T}&=&\sum_{i=1}^{n_1}\frac{\partial^2\log
f(\bfx_i,\bftheta_0)}{\partial\bftheta\partial\bftheta^T};
\\
\frac{\partial \bfq_{1n_2}(\bftheta_0,{\bf
0})}{\partial\bflambda^T}&=&-\sum_{i=1}^{n_2}\frac{\partial
\bfg(\bfy_i,\bftheta_0)}{\partial\bftheta^T};\\
\frac{\partial \bfq_{2n_2}(\bftheta_0,{\bf 0})}{\partial\bftheta^T}
&=&\sum_{i=1}^{n_2}\frac{\partial
\bfg(\bfy_i,\bftheta_0)}{\partial\bftheta^T};\\
\frac{\partial\bfq_{2n_2}(\bftheta_0,{\bf
0})}{\partial\bflambda^T}&=&-\sum_{i=1}^{n_2}
\bfg(\bfy_i,\bftheta_0)\bfg^T(\bfy_i,\bftheta_0).\\
\eeaa Hence we have
\[\renewcommand{\arraystretch}{0.6}\begin{pmatrix}
   -\displaystyle\frac{\partial
\bfw_{n_1}(\bftheta_0)}{\partial\bftheta^T}  &
-\displaystyle\frac{\partial \bfq_{1n_2}(\bftheta_0,{\bf
0})}{\partial\bflambda^T} \\ & \\
\displaystyle\frac{\partial \bfq_{2n_2}(\bftheta_0,{\bf
0})}{\partial\bftheta^T}    & \displaystyle\frac{\partial
    \bfq_{2n_2}(\bftheta_0,{\bf
0})}{\partial\bflambda^T} \end{pmatrix}
\begin{pmatrix}\hbftheta-\bftheta_0\\ \\
\hbflambda\end{pmatrix}=\begin{pmatrix}\bfw_{n_1}(\bftheta_0)+o_p(\bfdelta_{n})\\
\\ -\bfq_{2n_2}(\bftheta_0,{\bf
0})+o_p(\bfdelta_{n})\end{pmatrix},
\]
which can be written as \[\renewcommand{\arraystretch}{0.6}
\left(\begin{array}{c}\sqrt{n}(\hbftheta-\bftheta_0)\\ \\
\sqrt{n}\hbflambda\end{array}\right)=
V_n^{-1}\left(\begin{array}{c}\displaystyle\frac{1}{\sqrt{n}}\bfw_{n_1}(\bftheta_0)+
o_p\left(\displaystyle\frac{\bfdelta_{n}}{\sqrt{n}}\right)\\ \\
-\displaystyle\frac{1}{\sqrt{n}}\bfq_{2n_2}(\bftheta_0,{\bf
0})+o_p\left(\displaystyle\frac{\bfdelta_{n}}{\sqrt{n}}\right)\end{array}\right),
\]
where
\[\renewcommand{\arraystretch}{0.6} V_n=\left(\begin{array}{cc}
   -\displaystyle\frac{1}{n}\displaystyle\frac{\partial
\bfw_{n_1}(\bftheta_0)}{\partial\bftheta^T}
&-\displaystyle\frac{1}{n} \displaystyle\frac{\partial
\bfq_{1n_2}(\bftheta_0,{\bf 0})}{\partial\bflambda^T} \\ &\\
\displaystyle\frac{1}{n}
 \displaystyle\frac{\partial
\bfq_{2n_2}(\bftheta_0,{\bf 0})}{\partial\bftheta^T}
&\displaystyle\frac{1}{n} \displaystyle\frac{\partial
    \bfq_{2n_2}(\bftheta_0,{\bf
0})}{\partial\bflambda^T} \end{array}
  \right).\]
It can be shown that
\begin{equation} \nonumber
\begin{array}{l}
\bfdelta_{n}=O_p(n^{-1/2}); \\
{n}^{-1/2}\bfw_{n_1}(\bftheta_0)\stackrel{d}{\longrightarrow} N({\bf
0}, bI(\bftheta_0));\\
-{n}^{-1/2}\bfq_{2n_2}(\bftheta_0,{\bf
0})\stackrel{d}{\longrightarrow} N\Big({\bf
0},\;(1-b)E\big[\bfg(\bfY,\bftheta_0)\bfg^T(\bfY,\bftheta_0)\big]\Big);\\
\displaystyle\frac{1}{n}\displaystyle\frac{\partial
\bfw_{n_1}(\bftheta_0)}{\partial\bftheta^T}\stackrel{
\mbox{a.s.}}{\longrightarrow} -bI(\bftheta_0);\\
\displaystyle\frac{1}{{n}}\displaystyle\frac{\partial
\bfq_{1n_2}(\bftheta_0,{\bf 0})}{\partial\bflambda^T}\stackrel{
\mbox{a.s.}}{\longrightarrow}
-(1-b)E\left(\displaystyle\frac{\partial
\bfg(\bfY,\bftheta_0)}{\partial\bftheta^T}\right);\\
\displaystyle\frac{1}{n}\displaystyle\frac{\partial
\bfq_{2n_2}(\bftheta_0,{\bf
0})}{\partial\bftheta^T}\stackrel{\mbox{a.s.}}{\longrightarrow}
(1-b)E\left(\displaystyle\frac{\partial
\bfg(\bfY,\bftheta_0)}{\partial\bftheta^T}\right);\\
\displaystyle\frac{1}{n}\displaystyle\frac{\partial
    \bfq_{2n_2}(\bftheta_0,{\bf
0})}{\partial\bflambda^T}\stackrel{\mbox{a.s.}}{\longrightarrow}
-(1-b)E\big(\bfg(\bfY,\bftheta_0)\bfg^T(\bfY,\bftheta_0)\big).\\
\end{array}
\end{equation}
Therefore, it follows that
\[ \renewcommand{\arraystretch}{0.6}
\left(\begin{array}{c}\sqrt{n}(\hbftheta-\bftheta_0)\\
\sqrt{n}\hbflambda\end{array}\right)\stackrel{d}{\longrightarrow}N({\bf
0},W),\] where $W$ is the covariance matrix of
\[ \renewcommand{\arraystretch}{0.6}
V^{-1}\left(\begin{array}{c}\bfU_1\\
\bfU_2\end{array}\right)
\]
with
\[V=\left(\begin{array}{cc}
   bI(\bftheta_0) &
(1-b)E\left[\displaystyle\frac{\partial
\bfg(\bfY,\bftheta_0)}{\partial\bftheta^T}\right]^T\\
 (1-b)E\left[\displaystyle\frac{\partial
\bfg(\bfY,\bftheta_0)}{\partial\bftheta^T}\right]   &
-(1-b)E\big[\bfg(\bfY,\bftheta_0)\bfg^T(\bfY,\bftheta_0))\big]\end{array}
  \right)\equiv
  \begin{pmatrix}
            V_{11} & V_{12} \\
            V_{21} & V_{22} \\
          \end{pmatrix}
  ,\]
$\bfU_1$ is independent of $\bfU_2$ and
\[
\bfU_1\sim N({\bf 0}, bI(\bftheta_0)),\quad \bfU_2\sim N\Big({\bf
0},\;(1-b)E\big[\bfg(\bfY,\bftheta_0)\bfg^T(\bfY,\bftheta_0)\big]\Big).
\]
Since \bea \non V^{-1}&=&\left(\begin{array} {cc}
           V_{11}^{-1}+V_{11}^{-1}V_{12}V_{22.1}^{-1}V_{21}V_{11}^{-1} &
           -V_{22.1}^{-1}V_{21}V_{11}^{-1} \\  &\\
           -V_{11}^{-1}V_{12}V_{22.1}^{-1}& V_{22.1}^{-1} \\
         \end{array} \right)\\ \non &&\\ \non &&\\ \non
         &=&\begin{pmatrix}
           V_{11.2}^{-1} &
           -V_{11.2}^{-1}V_{12}V_{22}^{-1} \\  &\\
           -V_{22}^{-1}V_{21}V_{11.2}^{-1}& V_{22}^{-1}+
           V_{22}^{-1}V_{21}V_{11.2}^{-1}V_{12}V_{22}^{-1} \\
         \end{pmatrix}\equiv\begin{pmatrix}
            V^{(11)} & V^{(12)} \\
            V^{(21)} & V^{(22)} \\
          \end{pmatrix} \\ \non
\eea with $V_{11.2}=V_{11}-V_{12}V_{22}^{-1}V_{21}$ and
$V_{22.1}=V_{22}-V_{21}V_{11}^{-1}V_{12}$, it follows that \beaa
\sqrt{n}(\hbftheta-\bftheta_0) &\stackrel{d}{\longrightarrow}&N({\bf 0},S_1),\\
\sqrt{n}\hbflambda &\stackrel{d}{\longrightarrow}&N({\bf
0},S_2),\eeaa where \bea
S_1&=&V^{(11)}\cov(\bfU_1)(V^{(11)})^T+V^{(12)}\cov(\bfU_2)(V^{(12)})^T,\non\\
S_2&=&V^{(21)}\cov(\bfU_1)(V^{(21)})^T+V^{(22)}\cov(\bfU_2)(V^{(22)})^T.\hspace{4cm}
\square\non\eea

\subsection*{Proof of Theorem \ref{WEL thm: hypothesis}}

Assume that the assumptions made in Theorem \ref{WEL thm:
hypothesis} hold. A statistic for testing $H_0:\;
\bftheta=\bftheta_0$ is given by

\bea \non
 &&2[\ell(\hbftheta,\hbflambda)-\ell(\bftheta_0,\bflambda_0)]
\\
&=&2\left\{\sum_{i=1}^{n_1} \log
f(\bfx_i,\hbftheta)-\sum_{i=1}^{n_2}
\log\Big[1+\hbflambda^T\bfg({\bfy}_i,\hbftheta)\Big]\right\}\non\\
\label{RWEL: eqn thm 3.1}
 &&-2\left\{\sum_{i=1}^{n_2} \log
f(\bfx_i,\bftheta_0)-\sum_{i=1}^{n_2}
\log\Big[1+\bflambda_0^T\bfg({\bfy}_i,\bftheta_0)\Big]\right\}.
 \eea
Expending $\ell(\hbftheta,\hbflambda)$ around $(\bftheta_0,\zero)$
by Taylor's expansion, we have
 \bea \non && \ell(\hbftheta,\hbflambda)\\
 &=& \sum_{i=1}^{n_1} \log
f(\bfx_i,\hbftheta)-\sum_{i=1}^{n_2}
\log\Big[1+\hbflambda^T\bfg({\bfy}_i,\hbftheta)\Big]\non\\
&=&\sum_{i=1}^{n_1} \log f(\bfx_i,\bftheta_0)+
(\hbftheta-\bftheta_0)^T\bfw_{n_1}(\bftheta_0)+\displaystyle
\frac{1}{2}(\hbftheta-\bftheta_0)^T\frac{\partial
\bfw_{{n_1}}(\bftheta_0)}{\partial\bftheta^T}(\hbftheta-\bftheta_0)\non\\
&&-\hbflambda^T\bfq_{2n_2}(\bftheta_0,{\bf 0})-
\displaystyle\frac{1}{2}\hbflambda^T\frac{\partial\bfq_{2n_2}(\bftheta_0,{\bf
0})}{\partial\bflambda^T}\hbflambda
+\bfo_p(1)\non\\
&=&\sum_{i=1}^{n_1} \log f(\bfx_i,\bftheta_0)+
\left(\begin{array}{c}\displaystyle\frac{1}{\sqrt{n}}\bfw_{n_1}(\bftheta_0)\\
\\
 -\displaystyle\frac{1}{\sqrt{n}}\bfq_{2n_2}(\bftheta_0,{\bf
0})\end{array}\right)^TV^{-1}
\left(\begin{array}{c}\displaystyle\frac{1}{\sqrt{n}}\bfw_{n_1}(\bftheta_0)\\
\\
 -\displaystyle\frac{1}{\sqrt{n}}\bfq_{2n_2}(\bftheta_0,{\bf
0})\end{array}\right)\non\\
&&- \displaystyle\frac{1}{2}\left(\begin{array}{c}
\displaystyle\frac{1}{\sqrt{n}}\bfw_{n_1}(\bftheta_0)\\ \\
-\displaystyle\frac{1}{\sqrt{n}}\bfq_{2n_2}(\bftheta_0,{\bf
0})\end{array}\right)^T V^{-1}{\rm diag}(V_{11}\ V_{22})V^{-1}\left(\begin{array}{c}\displaystyle\frac{1}{\sqrt{n}}\bfw_{n_1}(\bftheta_0)\\ \\
-\displaystyle\frac{1}{\sqrt{n}}\bfq_{2n_2}(\bftheta_0,{\bf
0})\end{array}\right)\non\\
&&+\bfo_p(1). \label{RWEL: eqn thm 3.2}
 \eea
Under $H_0$, we have
\[ \displaystyle\frac{1}{n_2}\sum_{i=1}^{n_2}
\displaystyle\frac{\bfg({\bfy}_i,\bftheta_0)}
{\Big[1+\bflambda_0^T\bfg({\bfy}_i,\bftheta_0)\Big]}={\bf 0},\]
which implies that \beaa
\bflambda_0&=&\left[\displaystyle\frac{1}{n_2}\sum_{i=1}^{n_2}\bfg({\bfy}_i,\bftheta_0)
\bfg^T({\bfy}_i,\bftheta_0)\right]^{-1}
\left[\displaystyle\frac{1}{n_2}\sum_{i=1}^{n_2}\bfg({\bfy}_i,\bftheta_0)\right]+\bfo_p(1)
\\
&=& {-(1-b)}V_{22}^{-1}\displaystyle\frac{1}{n_2}
\bfq_{2n_2}(\bftheta_0,{\bf 0}) +\bfo_p(1).\eeaa Hence, \bea
&&\sum_{i=1}^{n_2}
\log\Big[1+\bflambda_0^T\bfg({\bfy}_i,\bftheta_0)\Big]\non\\
&=&\bflambda_0^T\sum_{i=1}^{n_2}\bfg({\bfy}_i,\bftheta_0)-
\displaystyle\frac{1}{2}\bflambda_0^T\sum_{i=1}^{n_2}\bfg({\bfy}_i,\bftheta_0)
\bfg^T({\bfy}_i,\bftheta_0)\bflambda_0 +\bfo_p(1)\non\\
&=&-\displaystyle\frac{n_2(1-b)}{2}\left[\displaystyle\frac{1}{n_2}\sum_{i=1}^{n_2}\bfg({\bfy}_i,\bftheta_0)\right]^T
V_{22}^{-1}\left[\displaystyle\frac{1}{n_2}\sum_{i=1}^{n_2}\bfg({\bfy}_i,\bftheta_0)\right]
+\bfo_p(1)\non\\ \label{RWEL: eqn thm 3.3}
&=&\displaystyle\frac{1}{2} \left[\displaystyle\frac{1}{\sqrt{n}}
\bfq_{2n_2}(\bftheta_0,{\bf 0})\right]^T
[-V_{22}]^{-1}\left[\displaystyle\frac{1}{\sqrt{n}}
\bfq_{2n_2}(\bftheta_0,{\bf 0})\right] +\bfo_p(1).\eea Substitute
equations (\ref{RWEL: eqn thm 3.2}) and (\ref{RWEL: eqn thm 3.3})
into equation (\ref{RWEL: eqn thm 3.1}), we have \bea
&&2\ell(\hbftheta,\hbflambda)-2\ell(\bftheta_0,\bflambda_0)\non\\
&=&2\left(\begin{array}{c}\displaystyle\frac{1}{\sqrt{n}}\bfw_{n_1}(\bftheta_0)\\
 \\ -\displaystyle\frac{1}{\sqrt{n}}\bfq_{2n_2}(\bftheta_0,{\bf
0})\end{array}\right)^TV^{-1}
\left(\begin{array}{c}\displaystyle\frac{1}{\sqrt{n}}\bfw_{n_1}(\bftheta_0)\\
\\
- \displaystyle\frac{1}{\sqrt{n}}\bfq_{2n_2}(\bftheta_0,{\bf
0})\end{array}\right)\non\\
&&-\left(\begin{array}{c}
\displaystyle\frac{1}{\sqrt{n}}\bfw_{n_1}(\bftheta_0)\\ \\
-\displaystyle\frac{1}{\sqrt{n}}\bfq_{2n_2}(\bftheta_0,{\bf
0})\end{array}\right)^T V^{-1}{\rm diag}(V_{11}\
V_{22})V^{-1}\left(\begin{array}{c}
\displaystyle\frac{1}{\sqrt{n}}\bfw_{n_1}(\bftheta_0)\\ \\
-\displaystyle\frac{1}{\sqrt{n}}\bfq_{2n_2}(\bftheta_0,{\bf
0})\end{array}\right)\non\\
&&+ \left[\displaystyle\frac{1}{\sqrt{n}}
\bfq_{2n_2}(\bftheta_0,{\bf 0})\right]^T
V_{22}^{-1}\left[\displaystyle\frac{1}{\sqrt{n}}
\bfq_{2n_2}(\bftheta_0,{\bf 0})\right]+\bfo_p(1)  \non \eea Hence
the limiting distribution of
$2[\ell(\hbftheta,\hbflambda)-\ell(\bftheta_0,\bflambda_0)]$ is the
same as the distribution of $
\left(\begin{array}{c}U_1\\
U_2\end{array}\right)^TV^{-1}\begin{pmatrix}
            V_{11.2} & V_{12} \\
           V_{21} & 0 \\
          \end{pmatrix}V^{-1}
           \left(\begin{array}{c}U_1\\
U_2\end{array}\right)$ under $H_0$. \\* \begin{flushright}
$\square$\end{flushright}

\clearpage

\begin{table} [h]
  \centering
  \caption{MSE Ratio of RSPELE to MLE with the smoothing parameter $h=n_2$ to the power of -1, -3/4, -1/2 and -1/4.}
  \renewcommand{\arraystretch}{0.75}
  \renewcommand{\tabcolsep}{0.4mm}
  \footnotesize \centering
  \begin{tabular}{ccccccccccc}
    \hline
    \hline \\
    &\multicolumn{10}{c}{\centering{\itshape Distribution of sample 2}}
    \\ \\
    n2 & $N(0,1)$ & $N(0,1.25)$ & $N(0,1.5)$ & $N(0,2)$ & $N(0,3)$ & \hspace{1.5em}$t_3$\hspace{1.5em}
     & \hspace{1.5em}$t_5$\hspace{1.5em} & $DE(0,0.5)$  & $DE(0,1)$ & $DE(0,1.5)$
     \\ \\
    \cline{2-11} \\
    &\multicolumn{10}{c}{\centering{\itshape $h=n_2^{-1}$}} \\ \\
    10 & 0.739 &  0.841 &  0.922  & 1.019 &  1.039 &  0.747 &  0.719 &  0.364 & 0.664 &  0.847  \\
    20 & 0.539 &  0.658 &  0.759  & 0.863 &  0.955 &  0.609 &  0.560 &  0.187 & 0.487 &  0.697  \\
    30 & 0.432 &  0.555 &  0.649  & 0.770 &  0.922 &  0.442 &  0.437 &  0.120 & 0.362 &  0.572  \\\\
    &\multicolumn{10}{c}{\centering{\itshape $h=n_2^{-3/4}$}} \\ \\
    10 & 0.700 &  0.810 &  0.878  & 0.987 &  1.036 &  0.730 &  0.692 &  0.339 & 0.645 &  0.821  \\
    20 & 0.503 &  0.624 &  0.734  & 0.853 &  0.941 &  0.588 &  0.512 &  0.172 & 0.466 &  0.662  \\
    30 & 0.405 &  0.528 &  0.624  & 0.765 &  0.892 &  0.418 &  0.413 &  0.107 & 0.333 &  0.552  \\\\
    &\multicolumn{10}{c}{\centering{\itshape $h=n_2^{-1/2}$}} \\ \\
    10 & 0.666 &  0.780 &  0.856  & 0.959 &  1.038 &  0.702 &  0.665 &  0.326 & 0.625 &  0.795  \\
    20 & 0.453 &  0.579 &  0.674  & 0.802 &  0.916 &  0.543 &  0.475 &  0.162 & 0.433 &  0.616  \\
    30 & 0.356 &  0.476 &  0.580  & 0.726 &  0.855 &  0.385 &  0.369 &  0.101 & 0.319 &  0.524  \\\\
    &\multicolumn{10}{c}{\centering{\itshape $h=n_2^{-1/4}$}} \\ \\
    10 & 0.616 &  0.752 &  0.842  & 0.943 &  1.036 &  0.664 &  0.630 &  0.329 & 0.630 &  0.779  \\
    20 & 0.403 &  0.532 &  0.635  & 0.773 &  0.899 &  0.504 &  0.438 &  0.174 & 0.419 &  0.600  \\
    30 & 0.320 &  0.434 &  0.532  & 0.682 &  0.838 &  0.356 &  0.338 &  0.111 & 0.315 &  0.506  \\
    \hline
    \hline
  \end{tabular}
  \label{RWEL table: Smooth with h=n_2}
\end{table}

\begin{table} [h]
  \centering
  \small
  \caption{Coverage Probability of MLE with the distribution of second population
  Normal, $t_3$, $t_5$ and Double Exponential, numbers in the brackets are
  AL, $\alpha=$ nominal level.}
  \renewcommand{\arraystretch}{0.75}
  \footnotesize \centering
  \setlength{\tabcolsep}{1pt}
  \begin{tabular}{ccccccccccc}
    \hline
    \hline
   $\alpha$ &\hspace{1.5em} &  0.800  &  0.900  &  0.950  &  0.990  &\hspace{1.5em} &  0.800  &  0.900  &  0.950  &  0.990 \\

   \hline\\
      $n_2$&& \multicolumn{4}{c}{\centering{Normal}}&&\multicolumn{4}{c}{\centering{$t_3$}}
      \\
           10 &&  0.774  &   0.864  &   0.913  &   0.972  & &   0.774  &   0.854  &   0.915  &   0.973   \\
              && (0.774) &  (0.994) &  (1.184) &  (1.557) & &  (0.777) &  (0.998) &  (1.189) &  (1.563)
              \\ \\
           20 &&  0.773  &   0.871  &   0.913  &   0.976  & &   0.769  &   0.849  &   0.901  &   0.964   \\
              && (0.784) &  (1.007) &  (1.200) &  (1.577) & &  (0.777) &  (0.997) &  (1.188) &  (1.562)  \\\\
           30 &&  0.810  &   0.888  &   0.932  &   0.983  & &   0.760  &   0.858  &   0.912  &   0.972   \\
              && (0.788) &  (1.011) &  (1.204) &  (1.583) & &  (0.786) &  (1.009) &  (1.203) &  (1.581)  \\\\

        && \multicolumn{4}{c}{\centering{$t_5$}}&&\multicolumn{4}{c}{\centering{Double Exponential}} \\
           10 &&  0.737  &   0.858  &   0.913  &   0.966  & &   0.771  &   0.863  &   0.913  &   0.962   \\
              && (0.789) &  (1.013) &  (1.207) &  (1.586) & &  (0.781) &  (1.002) &  (1.194) &  (1.569)  \\\\
           20 &&  0.759  &   0.862  &   0.915  &   0.975  & &   0.771  &   0.863  &   0.913  &   0.962   \\
              && (0.795) &  (1.020) &  (1.216) &  (1.598) & &  (0.781) &  (1.002) &  (1.194) &  (1.569)  \\\\
           30 &&  0.769  &   0.880  &   0.928  &   0.977  & &   0.771  &   0.863  &   0.913  &   0.962   \\
              && (0.790) &  (1.015) &  (1.209) &  (1.589) & &  (0.781) &  (1.002) &  (1.194) &  (1.569)  \\

    \hline
    \hline
  \end{tabular}
  \label{RWEL table: CP MLE}
\end{table}

\begin{table}[h]
  \centering
  \small
  \caption{Coverage Probability of RSPELE with different distributions of second population
  by using median estimating equation, numbers in the brackets are
  AL, $\alpha=$ nominal level.}
  \renewcommand{\arraystretch}{0.75}
  \footnotesize \centering
  \setlength{\tabcolsep}{0.35pt}
  \begin{tabular}{cccccccccccccccc}
    \hline
    \hline
   $\alpha$ &\hspace{3em}&  0.800  &  0.900  &  0.950  &  0.990  &\hspace{1.5em} &  0.800  &  0.900  &  0.950  &  0.990&\hspace{1.5em} &  0.800  &  0.900  &  0.950  &  0.990 \\

   \hline\\\\
      $n_2$&& \multicolumn{4}{c}{\centering{$N(0,1)$}}&&\multicolumn{4}{c}{\centering{$N(0,2)$}}&&\multicolumn{4}{c}{\centering{$DE(0,0.5)$}} \\
           10 &&  0.754    &  0.846    &  0.902    &  0.963    &&  0.746    &  0.839    &  0.898    &  0.960    &&  0.759    &  0.854    &  0.909    &  0.951  \\
              && (0.725)   & (0.930)   & (1.102)   & (1.435)   && (0.744)   & (0.949)   & (1.126)   & (1.471)   && (0.716)   & (0.919)   & (1.094)   & (1.441) \\\\
           20 &&  0.771    &  0.868    &  0.913    &  0.964    &&  0.756    &  0.855    &  0.908    &  0.958    &&  0.780    &  0.872    &  0.918    &  0.966  \\
              && (0.708)   & (0.910)   & (1.085)   & (1.415)   && (0.754)   & (0.961)   & (1.138)   & (1.482)   && (0.678)   & (0.878)   & (1.055)   & (1.401) \\\\
           30 &&  0.805    &  0.888    &  0.935    &  0.976    &&  0.791    &  0.873    &  0.925    &  0.970    &&  0.780    &  0.883    &  0.920    &  0.971  \\
              && (0.679)   & (0.874)   & (1.047)   & (1.389)   && (0.754)   & (0.957)   & (1.135)   & (1.483)   && (0.643)   & (0.843)   & (1.018)   & (1.368) \\\\

        && \multicolumn{4}{c}{\centering{$N(0,1.5)$}}&&\multicolumn{4}{c}{\centering{$t_3$}}&&\multicolumn{4}{c}{\centering{$DE(0,1)$}} \\\\
           10 &&  0.748    &  0.845    &  0.899    &  0.962    &&  0.758    &  0.848    &  0.908    &  0.964    &&  0.758    &  0.848    &  0.907    &  0.951  \\
              && (0.735)   & (0.945)   & (1.116)   & (1.457)   && (0.733)   & (0.938)   & (1.112)   & (1.460)   && (0.736)   & (0.941)   & (1.119)   & (1.460) \\\\
           20 &&  0.761    &  0.864    &  0.911    &  0.961    &&  0.766    &  0.851    &  0.892    &  0.951    &&  0.770    &  0.859    &  0.912    &  0.962  \\
              && (0.731)   & (0.938)   & (1.111)   & (1.447)   && (0.711)   & (0.911)   & (1.080)   & (1.411)   && (0.711)   & (0.911)   & (1.086)   & (1.425) \\\\
           30 &&  0.799    &  0.882    &  0.936    &  0.973    &&  0.771    &  0.869    &  0.916    &  0.966    &&  0.772    &  0.866    &  0.915    &  0.965  \\
              && (0.715)   & (0.912)   & (1.084)   & (1.422)   && (0.700)   & (0.897)   & (1.068)   & (1.402)   && (0.685)   & (0.882)   & (1.053)   & (1.392) \\\\
        && \multicolumn{4}{c}{\centering{$N(0,2)$}}&&\multicolumn{4}{c}{\centering{$t_5$}}&&\multicolumn{4}{c}{\centering{$DE(0,2)$}} \\\\
           10 &&  0.748    &  0.842    &  0.899    &  0.960    &&  0.734    &  0.840    &  0.897    &  0.963    &&  0.755    &  0.847    &  0.904    &  0.951  \\
              && (0.741)   & (0.948)   & (1.120)   & (1.464)   && (0.744)   & (0.949)   & (1.124)   & (1.467)   && (0.743)   & (0.952)   & (1.132)   & (1.470) \\\\
           20 &&  0.753    &  0.858    &  0.909    &  0.958    &&  0.761    &  0.858    &  0.906    &  0.974    &&  0.762    &  0.859    &  0.908    &  0.963  \\
              && (0.743)   & (0.951)   & (1.127)   & (1.462)   && (0.716)   & (0.916)   & (1.096)   & (1.441)   && (0.731)   & (0.935)   & (1.109)   & (1.445) \\\\
           30 &&  0.798    &  0.879    &  0.931    &  0.972    &&  0.766    &  0.877    &  0.929    &  0.971    &&  0.770    &  0.866    &  0.906    &  0.961  \\
              && (0.736)   & (0.936)   & (1.111)   & (1.456)   && (0.693)   & (0.892)   & (1.065)   & (1.403)   && (0.712)   & (0.913)   & (1.086)   & (1.416) \\

    \hline
    \hline
  \end{tabular}
  \label{RWEL table: CP RSPEL Median}
\end{table}

\begin{table}
  \centering
  \small
  \caption{Coverage Probability of RSPELE with different distributions of second population,
  and the smoothing parameter $h=n_2^{-1}$, numbers in the brackets are
  AL, $\alpha=$ nominal level.}
  \renewcommand{\arraystretch}{0.75}
  \footnotesize \centering
  \setlength{\tabcolsep}{0.35pt}
  \begin{tabular}{cccccccccccccccc}
    \hline
    \hline
   $\alpha$ &\hspace{2em}&  0.800  &  0.900  &  0.950  &  0.990  &\hspace{1.5em} &  0.800  &  0.900  &  0.950  &  0.990&\hspace{1.5em} &  0.800  &  0.900  &  0.950  &  0.990 \\

   \hline\\
      $n_2$&& \multicolumn{4}{c}{\centering{$N(0,1)$}}&&\multicolumn{4}{c}{\centering{$N(0,2)$}}&&\multicolumn{4}{c}{\centering{$DE(0,0.5)$}} \\
           10 &&  0.763  &   0.861  &   0.920  &   0.973  &&   0.752  &   0.840  &   0.902  &   0.966  &&   0.787  &   0.889  &   0.934  &   0.979  \\
              && (0.687) &  (0.886) &  (1.058) &  (1.394) &&  (0.754) &  (0.963) &  (1.139) &  (1.483) &&  (0.626) &  (0.829) &  (1.009) &  (1.368) \\\\
           20 &&  0.797  &   0.875  &   0.915  &   0.968  &&   0.765  &   0.863  &   0.907  &   0.956  &&   0.813  &   0.903  &   0.941  &   0.979  \\
              && (0.649) &  (0.840) &  (1.006) &  (1.344) &&  (0.756) &  (0.963) &  (1.138) &  (1.477) &&  (0.568) &  (0.766) &  (0.945) &  (1.311) \\\\
           30 &&  0.816  &   0.912  &   0.941  &   0.980  &&   0.802  &   0.885  &   0.926  &   0.968  &&   0.821  &   0.901  &   0.946  &   0.979  \\
              && (0.610) &  (0.796) &  (0.965) &  (1.305) &&  (0.751) &  (0.956) &  (1.130) &  (1.478) &&  (0.523) &  (0.718) &  (0.892) &  (1.263) \\\\

        && \multicolumn{4}{c}{\centering{$N(0,1.5)$}}&&\multicolumn{4}{c}{\centering{$t_3$}}&&\multicolumn{4}{c}{\centering{$DE(0,1)$}} \\\\
           10 &&  0.756  &   0.855  &   0.914  &   0.972  &&   0.782  &   0.873  &   0.923  &   0.968  &&   0.777  &   0.869  &   0.920  &   0.965  \\
              && (0.726) &  (0.930) &  (1.104) &  (1.448) &&  (0.709) &  (0.910) &  (1.082) &  (1.441) &&  (0.703) &  (0.909) &  (1.084) &  (1.438) \\\\
           20 &&  0.786  &   0.873  &   0.913  &   0.966  &&   0.774  &   0.869  &   0.911  &   0.958  &&   0.794  &   0.875  &   0.927  &   0.970  \\
              && (0.707) &  (0.908) &  (1.077) &  (1.415) &&  (0.668) &  (0.861) &  (1.028) &  (1.361) &&  (0.660) &  (0.856) &  (1.029) &  (1.370) \\\\
           30 &&  0.803  &   0.899  &   0.935  &   0.974  &&   0.793  &   0.880  &   0.924  &   0.971  &&   0.785  &   0.878  &   0.926  &   0.971  \\
              && (0.684) &  (0.874) &  (1.041) &  (1.377) &&  (0.648) &  (0.836) &  (1.001) &  (1.341) &&  (0.622) &  (0.811) &  (0.981) &  (1.320) \\\\
        && \multicolumn{4}{c}{\centering{$N(0,2)$}}&&\multicolumn{4}{c}{\centering{$t_5$}}&&\multicolumn{4}{c}{\centering{$DE(0,2)$}} \\\\
           10 &&  0.752  &   0.846  &   0.910  &   0.970  &&   0.765  &   0.865  &   0.911  &   0.964  &&   0.774  &   0.862  &   0.915  &   0.958  \\
              && (0.743) &  (0.948) &  (1.123) &  (1.473) &&  (0.716) &  (0.914) &  (1.091) &  (1.436) &&  (0.732) &  (0.943) &  (1.121) &  (1.467) \\\\
           20 &&  0.771  &   0.871  &   0.909  &   0.961  &&   0.785  &   0.881  &   0.922  &   0.976  &&   0.780  &   0.868  &   0.919  &   0.966  \\
              && (0.732) &  (0.936) &  (1.112) &  (1.453) &&  (0.666) &  (0.855) &  (1.028) &  (1.376) &&  (0.707) &  (0.907) &  (1.078) &  (1.417) \\\\
           30 &&  0.798  &   0.893  &   0.934  &   0.973  &&   0.799  &   0.893  &   0.939  &   0.974  &&   0.780  &   0.874  &   0.917  &   0.965  \\
              && (0.720) &  (0.917) &  (1.087) &  (1.433) &&  (0.637) &  (0.822) &  (0.990) &  (1.334) &&  (0.677) &  (0.873) &  (1.043) &  (1.381) \\

    \hline
    \hline
  \end{tabular}
  \label{RWEL table: CP RSPEL n_2^{-1}}
\end{table}

\begin{table}
  \centering
  \small
  \caption{Coverage Probability of RSPELE with different distributions of second population,
  and the smoothing parameter $h=n_2^{-1/2}$, numbers in the brackets are
  AL, $\alpha=$ nominal level.}
  \renewcommand{\arraystretch}{0.75}
  \footnotesize \centering
  \setlength{\tabcolsep}{0.35pt}
  \begin{tabular}{ccccccccccccccc}
    \hline
    \hline
   $\alpha$ &  0.800  &  0.900  &  0.950  &  0.990  &\hspace{1.5em} &  0.800  &  0.900  &  0.950  &  0.990&\hspace{1.5em} &  0.800  &  0.900  &  0.950  &  0.990 \\

   \hline\\
      $n_2$& \multicolumn{4}{c}{\centering{$N(0,1)$}}&&\multicolumn{4}{c}{\centering{$N(0,2)$}}&&\multicolumn{4}{c}{\centering{$DE(0,0.5)$}} \\
           10 &  0.770  &   0.855  &   0.918  &   0.974  &&   0.750   &  0.846  &   0.908  &   0.970  &&   0.773  &   0.877  &   0.933  &   0.980    \\
              & (0.630) &  (0.814) &  (0.977) &  (1.310) &&  (0.756)  & (0.965) &  (1.143) &  (1.491) &&  (0.508) &  (0.685) &  (0.861) &  (1.240)   \\\\
           20 &  0.777  &   0.870  &   0.921  &   0.973  &&   0.773   &  0.865  &   0.902  &   0.954  &&   0.797  &   0.899  &   0.945  &   0.982    \\
              & (0.551) &  (0.713) &  (0.862) &  (1.176) &&  (0.745)  & (0.953) &  (1.127) &  (1.459) &&  (0.391) &  (0.531) &  (0.676) &  (1.051)   \\\\
           30 &  0.805  &   0.904  &   0.936  &   0.980  &&   0.800   &  0.887  &   0.933  &   0.972  &&   0.795  &   0.893  &   0.943  &   0.987    \\
              & (0.491) &  (0.636) &  (0.770) &  (1.079) &&  (0.732)  & (0.933) &  (1.106) &  (1.442) &&  (0.313) &  (0.431) &  (0.568) &  (0.910)   \\\\
        & \multicolumn{4}{c}{\centering{$N(0,1.5)$}}&&\multicolumn{4}{c}{\centering{$t_3$}}&&\multicolumn{4}{c}{\centering{$DE(0,1)$}} \\\\
           10 &  0.763  &   0.849  &   0.911  &   0.970  &&   0.782   &  0.874  &   0.923  &   0.966  &&   0.758  &   0.876  &   0.925  &   0.969    \\
              & (0.702) &  (0.903) &  (1.074) &  (1.416) &&  (0.662)  & (0.853) &  (1.021) &  (1.375) &&  (0.649) &  (0.845) &  (1.019) &  (1.377)   \\\\
           20 &  0.780  &   0.864  &   0.911  &   0.964  &&   0.782   &  0.862  &   0.916  &   0.967  &&   0.791  &   0.889  &   0.923  &   0.975    \\
              & (0.657) &  (0.844) &  (1.003) &  (1.327) &&  (0.589)  & (0.763) &  (0.918) &  (1.241) &&  (0.566) &  (0.740) &  (0.897) &  (1.235)   \\\\
           30 &  0.816  &   0.900  &   0.934  &   0.976  &&   0.797   &  0.881  &   0.928  &   0.978  &&   0.777  &   0.885  &   0.928  &   0.978    \\
              & (0.611) &  (0.787) &  (0.939) &  (1.258) &&  (0.538)  & (0.698) &  (0.845) &  (1.152) &&  (0.493) &  (0.649) &  (0.795) &  (1.123)   \\\\
        & \multicolumn{4}{c}{\centering{$N(0,2)$}}&&\multicolumn{4}{c}{\centering{$t_5$}}&&\multicolumn{4}{c}{\centering{$DE(0,2)$}} \\\\
           10 &  0.753  &   0.852  &   0.905  &   0.970  &&   0.766   &  0.862  &   0.920  &   0.967  &&   0.755  &   0.864  &   0.908  &   0.967    \\
              & (0.733) &  (0.940) &  (1.113) &  (1.461) &&  (0.662)  & (0.854) &  (1.021) &  (1.365) &&  (0.708) &  (0.910) &  (1.087) &  (1.446)   \\\\
           20 &  0.774  &   0.865  &   0.906  &   0.959  &&   0.786   &  0.880  &   0.936  &   0.972  &&   0.786  &   0.866  &   0.922  &   0.963    \\
              & (0.706) &  (0.906) &  (1.073) &  (1.402) &&  (0.578)  & (0.747) &  (0.900) &  (1.226) &&  (0.649) &  (0.836) &  (1.002) &  (1.342)   \\\\
           30 &  0.809  &   0.899  &   0.935  &   0.973  &&   0.803   &  0.901  &   0.940  &   0.975  &&   0.785  &   0.875  &   0.919  &   0.968    \\
              & (0.677) &  (0.862) &  (1.026) &  (1.359) &&  (0.520)  & (0.679) &  (0.823) &  (1.137) &&  (0.596) &  (0.774) &  (0.934) &  (1.264)   \\

    \hline
    \hline
  \end{tabular}
  \label{RWEL table: CP RSPEL n_2^{-1/2}}
\end{table}

\end{document}